\documentclass[twocolumn,showpacs,preprintnumbers,amsmath,amssymb,superscriptaddress]{revtex4}

\usepackage{graphicx}
\begin{document}
\bibliographystyle{prsty}
\title{Chemical potential shift induced by double-exchange and polaronic effects in Nd$_{1-x}$Sr$_x$MnO$_3$}

\author{K. Ebata}
\affiliation{Department of Physics and Department of Complexity Science and Engineering, University of Tokyo, 7-3-1 Hongo, Bunkyo-ku, Tokyo 113-0033, Japan}
\author{M. Takizawa}
\affiliation{Department of Physics and Department of Complexity Science and Engineering, University of Tokyo, 7-3-1 Hongo, Bunkyo-ku, Tokyo 113-0033, Japan}
\author{K. Maekawa}
\affiliation{Department of Physics and Department of Complexity Science and Engineering, University of Tokyo, 7-3-1 Hongo, Bunkyo-ku, Tokyo 113-0033, Japan}
\author{A. Fujimori}
\affiliation{Department of Physics and Department of Complexity Science and Engineering, University of Tokyo, 7-3-1 Hongo, Bunkyo-ku, Tokyo 113-0033, Japan}
\author{H. Kuwahara}
\affiliation{Department of Physics, Sophia University, Chiyoda-ku, Tokyo 102-8554, Japan}
\author{Y. Tomioka}
\affiliation{Correlated Electron Research Center (CERC), National Institute of Advanced Industrial Science and Technology (AIST), Tsukuba 305-8562, Japan}
\author{Y. Tokura}
\affiliation{Correlated Electron Research Center (CERC), National Institute of Advanced Industrial Science and Technology (AIST), Tsukuba 305-8562, Japan}
\affiliation{Department of Applied Physics, University of Tokyo, Bunkyo-ku, Tokyo 113-8656, Japan}
\affiliation{Spin Superstructure Project, Exploratory Research for Advanced Technology (ERATO), Japan Science and Technology Corporation (JST), Tsukuba 305-8562, Japan}
\date{\today}

\begin{abstract}
We have studied the chemical potential shift as a function of temperature in Nd$_{1-x}$Sr$_x$MnO$_3$ (NSMO) by measurements of core-level photoemission spectra.
For ferromagnetic samples ($x=0.4$ and 0.45), we observed an unusually large upward chemical potential shift with decreasing temperature in the low-temperature region of the ferromagnetic metallic (FM) phase. This can be explained by the double-exchange (DE) mechanism if the $e_g$ band is split by dynamical/local Jahn-Teller effect. The shift was suppressed near the Curie temperature ($T_C$), which we attribute to the crossover from the DE to lattice-polaron regimes.
\end{abstract}


\maketitle
\section{Introduction}
Perovskite-type manganites of the chemical formula $R_{1-x}A_x$MnO$_3$ (where $R$ is a rare-earth and $A$ is an alkaline-earth metal) have attracted much interest because of their colossal magnetoresistance (CMR) \cite{Tokura5}.
Historically, the magnetic and transport properties of this kind of materials have been understood in terms of double-exchange (DE) interaction between the localized $t_{2g}$ electrons and the itinerant $e_g$ electrons \cite{Zener, Anderson, Furukawa2}. In this model, the kinetic energy gain of the holes doped into the $e_g$ band is maximized when the spins of the $t_{2g}$ electrons are aligned in the same direction, thereby stabilizing the ferromagnetic ground state. However, it has been pointed out by Millis {\it et al.} \cite{Millis} that the DE model is insufficient to explain the huge change in the resistivity under a magnetic field and the high resistivity above the Curie temperature ($T_C$). They have proposed a model including the dynamical Jahn-Teller (JT) distortion of the MnO$_6$ octahedra in addition to the DE interaction and reproduced the experimentally observed behavior of the resistivity. It has also been proposed that the extraordinary enhancement of the resistivity in the manganites may result from the emergence of lattice polarons in the paramagnetic insulating (PI) phase \cite{shimomura, Kiryukhin, Dai, Millis}. Koo {\it et al.} \cite{Koo} have found that lattice polarons in Nd$_{0.7}$Sr$_{0.3}$MnO$_3$ are strongly suppressed by applying magnetic field and do not completely disappear at high fields, corresponding to an admixture of the ``conducting" and ``insulating" carriers.
From x-ray scattering and neutron scattering studies \cite{shimomura, Kiryukhin, Dai}, indeed, the diffuse scattering due to the lattice polarons has been observed in the high-temperature PI phase of manganites and gradually disappears with decreasing temperature in the ferromagnetic metallic (FM) phase. On the other hand, several local structural studies have suggested that a dynamical or local JT distortion persists in the FM phase, too. X-ray absorption fine structure (EXAFS) studies have shown that the lattice distortion of the MnO$_6$ octahedra is found in the FM phase of La$_{1-x}$Ca$_{x}$MnO$_3$ at low temperatures \cite{Lanzara, Booth}. From the pair-density function analysis of pulsed neutron diffraction data, the local JT distortion has been observed in the FM phase of La$_{1-x}$Sr$_{x}$MnO$_3$ \cite{Louca}. If the JT distortion exists in the FM phase, the degeneracy of the $e_g$ band may be lifted already in that phase.

\begin{figure}
\begin{center}
\includegraphics[width=7cm]{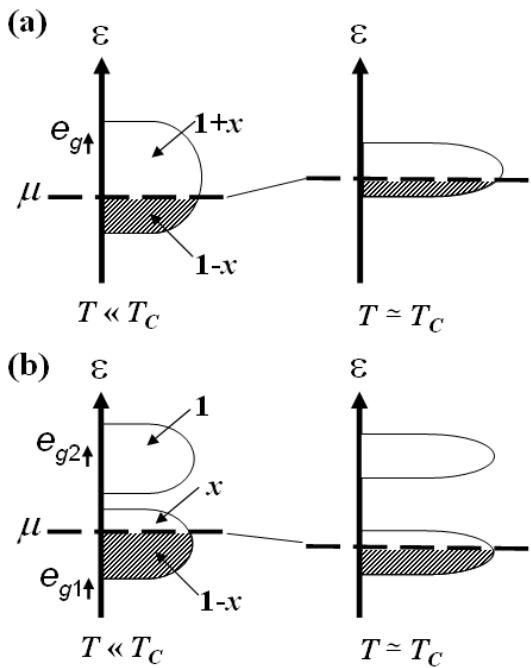}
\caption{Schematic pictures of the temperature-dependent DOS due to the DE interaction in FM phase for $x$$<$0.5. (a) Degenerate two-orbital model; (b) One-orbital model resulting from the Jahn-Teller splitting. $\mu$, $T_C$ and $x$ denote the chemical potential, the Curie temperature and the hole concentration, respectively.}
\label{DEmodel}
\end{center}
\end{figure}
In order to obtain insight into the competition between the DE mechanism and the lattice-polaron effect, the measurement of chemical potential shift ($\Delta \mu$) as a function of temperature gives much insight. According to the DE model, temperature-dependent chemical potential shift occurs due to the change in the $e_g$ band width as schematically shown in Fig. 1 \cite{Furukawa}.
If the $e_g$ band is split by JT distortion and the one-orbital DE model becomes relevant, the chemical potential is shifted upward for hole concentration $x$$<$0.5 with decreasing temperature and downward for $x$$>$0.5 \cite{Furukawa}. If one takes into account the double degeneracy of the $e_g$ orbitals [Fig. 1(a)], a downward shift with decreasing temperature would be expected in the FM phase for 0$<$$x$$<$1 because the the up-spin band of the $e_g$ orbitals is less than half-filled in the $R_{1-x}A_x$MnO$_3$ compounds. Therefore, the temperature-dependent chemical potential shift is sensitive to the splitting of the $e_g$ band and therefore to the dynamical/local JT effect. Schulte {\it et al.} \cite{Schulte} have investigated the change of the work function in La$_{1.2}$Sr$_{1.8}$Mn$_2$O$_3$ as a function of temperature by measurements of photoemission spectra and attributed the change to the temperature-dependent shift of chemical potential. Alternatively, the chemical potential shift can be deduced from the shifts of photoemission spectra because the binding energies of the spectra are measured relative to the chemical potential. We employ the latter method in this work.

\begin{figure}
\begin{center}
\includegraphics[width=6.5cm]{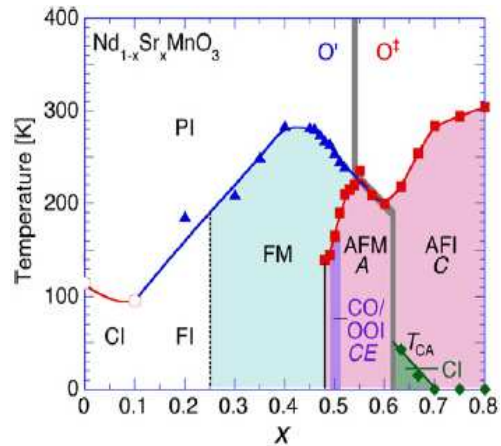}
\caption{(Color online) Electronic phase diagram of Nd$_{1-x}$Sr$_x$MnO$_3$. PI: Paramagnetic insulating phase; CI: Spin-canted insulating phase; FM: Ferromagnetic metallic phase; FI: Ferromagnetic insulating phase; CO/OOI: Charge-orbital ordered insulating phase; AFM: Antiferromagnetic metallic phase; AFI: Antiferromagnetic insulating phase \cite{Kuwahara, Tokura5}.}
\label{phase}
\end{center}
\end{figure}
Nd$_{1-x}$Sr$_x$MnO$_3$ (NSMO) is a suitable system for clarifying the relationship between the DE interactions and the existence of lattice polarons because it shows the FM phase for $x \alt 0.5$ and the so-called CE-type antiferromagnetic (AF) charge-ordered (CO) phase in the doping region close to the half-doping $x$ = 0.5 as shown in Fig. 2 \cite{Kuwahara, Tokura5}.
In this work, we study the chemical potential shift in NSMO as a function of temperature by measurements of core-level photoemission spectra. We found that the chemical potential shift was large in the low-temperature part of the FM phase as predicted by the DE model and dynamical/local JT effect and was suppressed at high temperatures near $T_C$. We consider that the different behaviors with temperature to be related to the competition between the DE interaction and the lattice-polaron effect.

\section{Experimental}
Single crystals of NSMO ($x=0.4$ and 0.45) were prepared by the floating zone method \cite{Kuwahara}.
X-ray photoemission spectroscopy measurements were performed using the photon energies of $h\nu =$ 1253.6 eV (Mg $K \alpha$). All the photoemission measurements were performed under the base pressure of $\sim 10^{-10}$ Torr at 20-330 K. The samples were repeatedly scraped {\it in situ} with a diamond file to obtain clean surfaces.
The cleanliness of the sample surface was checked by the reduction of the shoulder on the high binding energy side of the O 1$s$ core level.
Photoelectrons were collected using a Scienta SES-100 electron-energy analyzer. The energy resolution was about 800 meV. The measured binding energies were stable, because the gold $4{\it f}_{7/2}$ core-level spectrum did not changed in the measurements with the accuracy of $\pm 10$ meV at each temperature.

\section{Results and discussion}
In Fig. 3, we have plotted the spectra of the O $1s$, Sr $3d$, Nd $3d$ and Mn $2p$ core levels in NSMO with $x=0.4$. The vertical lines mark the estimated positions of the core levels used in the present study. We employed the midpoint of the low binding-energy slope for the O $1s$ core level because the line shape on the higher-binding energy side of the O $1s$ spectra is known to be affected by surface contamination or degradation.
We also employed the midpoint for the Sr $3d$ and Mn $2p$ core levels.
As for the Nd $3d$ core level, 80 $\%$ of the peak height of the low binding-energy slope was used because the line shape near the midpoint on the lower-binding energy side slightly changed with temperature.
\begin{figure}
\begin{center}
\includegraphics[width=9.3cm]{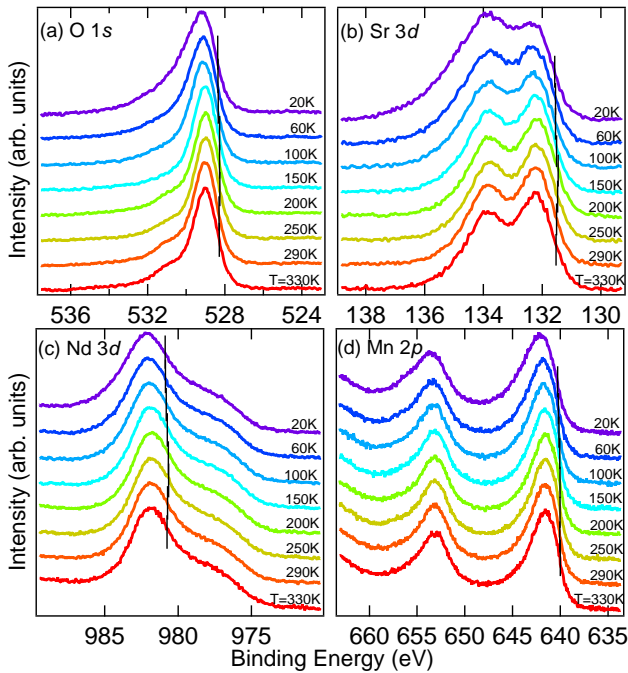}
\caption{(Color online) Core-level photoemission spectra of Nd$_{0.6}$Sr$_{0.4}$MnO$_3$ taken with the Mg $K \alpha$ line. (a) O $1s$; (b) Sr $3d$; (c) Nd $3d$; (d) Mn $2p$.}
\label{core1}
\end{center}
\end{figure}

Figure 4(a) shows the binding energy shift of each core level as a function of temperature for NSMO with $x=0.4$. One can see that the observed binding energy shifts with temperature were approximately common to the O $1s$, Sr $3d$, Nd $3d$ and Mn $2p$ core levels. From the measurements of the doping-dependent chemical potential shift in Pr$_{1-x}$Ca$_x$MnO$_3$, all the core levels have shown identical shifts with hole concentration except for the Mn $2p$ core level, where the effect of chemical shift is superimposed \cite{Ebata}. Here, all the core levels including the Mn 2$p$ core level exhibit similar shifts with temperature in contrast to the measurements of doping-dependent core-level shifts \cite{Ebata}. Therefore, we assume that the shifts of the core levels are largely due to the chemical potential shift, and take the average of the shifts of the four core levels as a measure of $\Delta \mu$ as a function of temperature in NSMO.

\begin{figure}
\begin{center}
\includegraphics[width=9cm]{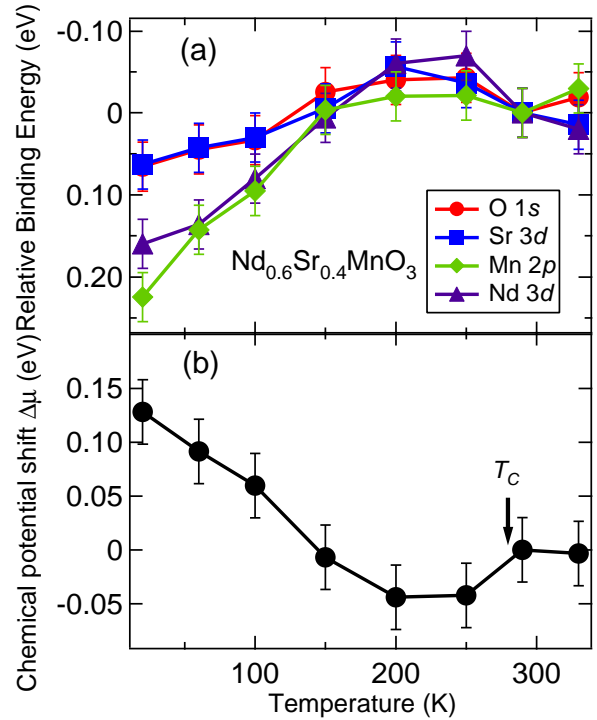}
\caption{(Color online) Core-level shifts and chemical potential shift in Nd$_{0.6}$Sr$_{0.4}$MnO$_3$. (a) Binding energy shifts of the O $1s$, Sr $3d$, Nd $3d$, and Mn $2p$ core levels as functions of temperature relative to 290 K; (b) Chemical potential shift $\Delta \mu$ as a function of temperature. $T_C$ denotes Curie temperature.}
\label{chemicalpotential}
\end{center}
\end{figure}
Figure 4 (b) shows the temperature-dependent chemical potential shift in NSMO with $x=0.4$. We observed a large upward chemical potential shift with decreasing temperature in the FM phase of NSMO at low temperatures. Furukawa \cite{Furukawa} proposed an anomalous temperature-dependent chemical potential shift below $T_C$ due to the DE interaction through the change of the $e_g$ band width with temperature. He also predicted that the magnitude of the shift was estimated to be about 0.1 eV when the $e_g$ band width was of order $\sim$ 1 eV \cite{Furukawa}. If the $e_g$ band remains degenerate in the FM phase of NSMO, one would expect to see a downward chemical potential shift with decreasing temperature based on the DE interaction because the up-spin band of the $e_g$ orbitals is less than half-filled for 0$<$$x$$<$1. On the other hand, if the degeneracy of the $e_g$ band is lifted by the dynamical/local JT distortion, the upward chemical potential shift with decreasing temperature is predicted because the band is more than half filled for $x$$<$0.5 (see Fig. 1). Therefore, we attribute the large upward shift of the chemical potential with decreasing temperature in the low-temperature FM phase of NSMO to the change in the width of the JT-split $e_g$ band caused by the DE mechanism. The magnitude of the observed shift is in quantitative agreement with the results of the one-orbital DE model \cite{Furukawa}.

We consider that the temperature-dependent splitting of the $e_g$ band may also influence the temperature-dependent chemical potential shift in the FM phase of NSMO because the intensity of the diffuse scattering due to JT effect gradually increases with increasing temperature in the FM phase, which indicates further increase of the energy splitting of the $e_g$ level \cite{shimomura, Kiryukhin, Dai}. If this is the case, the temperature-dependent shift would be even stronger at higher temperatures for the one-orbital case and the suppression of chemical potential shift in NSMO within the FM phase near $T_C$ cannot be explained. At high temperatures near $T_C$ in the PI phase, the diffuse scattering due to the formation of lattice polarons has been observed by means of x-ray scattering and neutron scattering studies \cite{shimomura, Kiryukhin, Dai}. We consider that the suppression of the shift in the FM phase at high temperatures is connected with the influence of the lattice polarons and the DE model is no more effective at those high temperatures.

\begin{figure}
\begin{center}
\includegraphics[width=9cm]{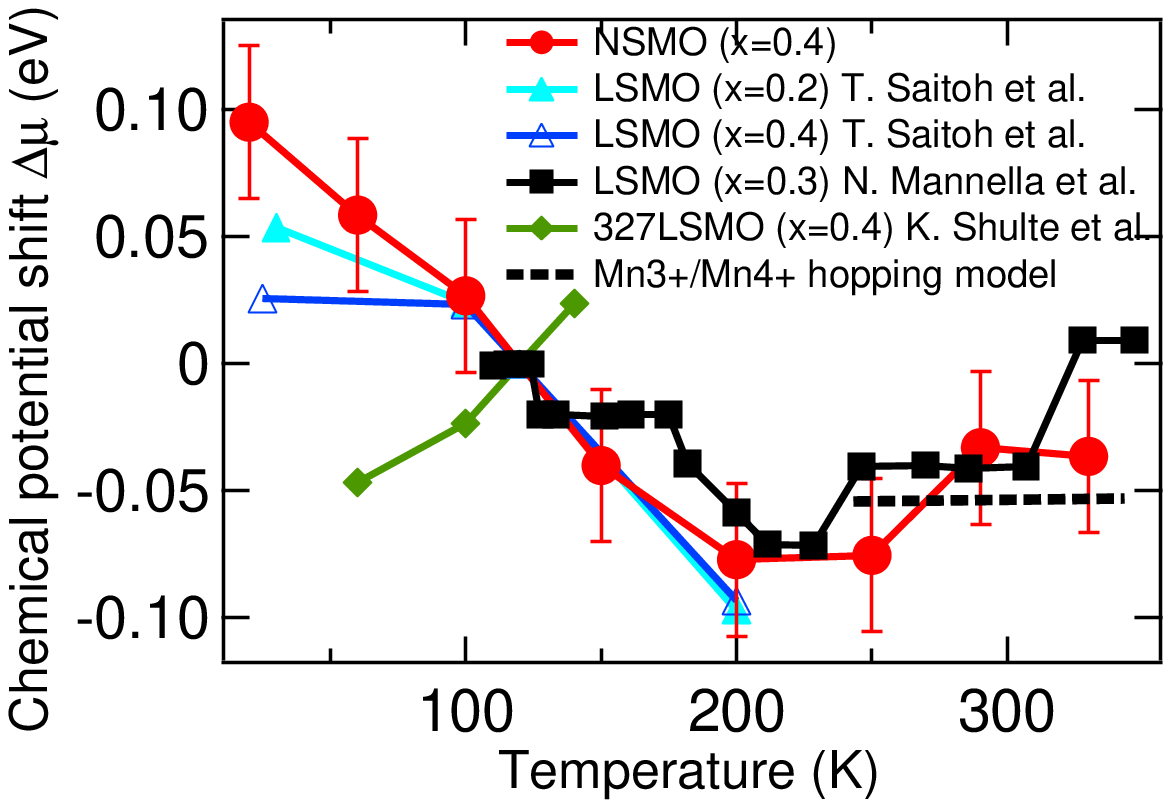}
\caption{(Color online) Comparison of the chemical potential shift $\Delta \mu$ in Nd$_{1-x}$Sr$_x$MnO$_3$ ($x=0.4$) with the shifts of the O 1$s$ core level (in the process of increasing temperature) and the valence band of La$_{1-x}$Sr$_x$MnO$_3$ ($x=0.2$, 0.3 and 0.4 with $T_C\approx310$, 370, and 370 K, respectively) \cite{Mannella, Saitoh}, $\Delta \mu$ for the bilayered system La$_{2-2x}$Sr$_{1+2x}$Mn$_2$O$_7$ ($x=0.4$ with $T_C\approx125$ K) \cite{Schulte} and correlated hopping model of Mn$^{3+}$/Mn$^{4+}$ mixed-valence state. The spectral shifts have been translated to the chemical potential shift, assuming that the spectral shifts are primarily close to the chemical potential shift.}
\label{comparison}
\end{center}
\end{figure}
In Fig. 5, we compare $\Delta \mu$ for NSMO ($x=0.4$) with the shifts of the O 1$s$ core level (in the process of increasing temperature) and the valence band of La$_{1-x}$Sr$_x$MnO$_3$ ($x=0.2$, 0.3 and 0.4), $\Delta \mu$ for bilayered system La$_{2-2x}$Sr$_{1+2x}$Mn$_2$O$_7$ ($x=0.4$) and correlated hopping model of Mn$^{3+}$/Mn$^{4+}$ mixed-valence state as a function of temperature \cite{Mannella, Saitoh, Schulte, Ishida, Chaikin, Koshibae}. We consider that the temperature-dependent shifts of the O 1$s$ core level and the valence band for La$_{1-x}$Sr$_x$MnO$_3$ are ascribed to the temperature-dependent change in $\Delta \mu$. For NSMO and La$_{1-x}$Sr$_x$MnO$_3$, the $\Delta \mu$ curves show similar temperature dependences in the sense that the chemical potential is shifted upward with decreasing temperature. We consider that the shifts are understood in terms of the DE interaction (and possibly the temperature dependence of the splitting of the $e_g$ band caused by the dynamical/local JT effect) \cite{Furukawa, shimomura, Kiryukhin, Dai}. An upward chemical potential shift by applying magnetic field has also been observed in La$_{1-x}$Sr$_x$MnO$_3$, which may be ascribed to the DE interaction in the JT split $e_g$ band \cite{Wu}. However, the $\Delta \mu$ for the bilayered compound La$_{2-2x}$Sr$_{1+2x}$Mn$_2$O$_7$ reported by Schulte {\it et al.} \cite{Schulte} is opposite to the prediction of the DE model.
They speculated that the two dimensionality might affect the temperature-dependent chemical potential shift, but its origin remains as an open question.

Also, we have compared the experimentally observed shifts with the results of the correlated hopping model in the splitting of the $e_g$ band due to the dynamical/local JT distortion \cite{Ishida, Chaikin, Koshibae}. The correlated hopping model is given by $\Delta\mu = -k_B\ln\bigl(\frac{g_3}
{g_4}\frac{x}{1-x}\bigr)\times T + {\rm const.}$,
where $g_3$ and $g_4$ are the spin-orbital degeneracies of Mn$^{3+}$ and Mn$^{4+}$, respectively and $x$ is the fraction of Mn$^{4+}$ ions \cite {Ishida, Chaikin, Koshibae}. The parameters were fixed at $g_3=5$, $g_4=4$ and $x=0.4$, corresponding to the Mn$^{3+}$/Mn$^{4+}$ mixed-valence state in the split of the $e_g$ band. We have plotted the calculated $\Delta \mu$ as shown by a dashed line in Fig. 5 \cite {Ishida, Chaikin, Koshibae}. The result is in qualitative agreement with the observed chemical potential shift in high-temperature region of NSMO with $x=0.4$.

\begin{figure}
\begin{center}
\includegraphics[width=9cm]{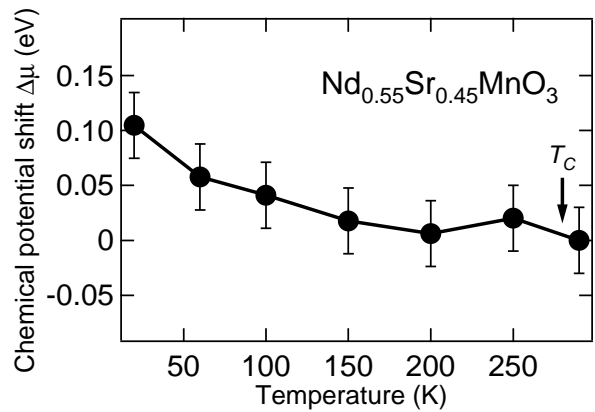}
\caption{Chemical potential shift in Nd$_{0.55}$Sr$_{0.45}$MnO$_3$. $T_C$ denotes Curie temperature.}
\label{shift_FM}
\end{center}
\end{figure}
For other hole concentrations of NSMO, too, we confirmed that the observed temperature-dependent shifts were common to the O $1s$, Sr $3d$, Nd $3d$ and Mn $2p$ core levels and deduced the temperature-dependent chemical potential shifts by taking the average of the shifts of the four core levels. Figure 6 shows the $\Delta \mu$ for NSMO with $x=0.45$ as a function of temperature. The shift $\Delta \mu$ was suppressed just below the $T_C$ as in the case of $x=0.4$, in accordance with the enhancement of the diffuse scattering due to the lattice polarons \cite{shimomura, Kiryukhin, Dai}. We attribute the shift in the low-temperature region of the FM phase for $x=0.45$ to the DE interaction (and the temperature dependence of the splitting of the $e_g$ band due to the dynamical/local JT effect) as in the case of $x=0.4$. \cite{Furukawa, shimomura, Kiryukhin, Dai}. The magnitude of the observed shift for $x=0.45$ is smaller than that for $x=0.4$, consistent with the one-orbital DE model resulting from the JT splitting [see Fig. 1(b)].

\section{Conclusion}
We have measured the chemical potential shift as a function of temperature in NSMO by means of core-level photoemission spectroscopy.
We have found an anomalous upward chemical potential shift with decreasing temperature in the low-temperature region of the FM phase and its suppression in the high-temperature region of the FM phase near $T_C$. We attribute the large shift in the low-temperature region to the change of band width due to the DE interaction (and possibly the temperature dependence of the splitting of the $e_g$ level caused by the dynamical/local JT effect.) Also, the suppression of the shift at higher temperatures is ascribed to the influence of lattice-polaron formation. 

\section{Acknowledgment}
Informative discussions with N. Furukawa and H. Wadati are gratefully acknowledged.
This work was supported by a Grant-in-Aid for Scientific Research in Priority Area ``Invention of Anomalous Quantum Materials" from the Ministry of Education, Culture, Sports, Science and Technology, Japan.


\begin{thebibliography}{10}

\bibitem{Tokura5}
Y. Tokura, Rep. Prog. Phys {\bf 69},  797  (2006).

\bibitem{Zener}
C. Zener, Phys. Rev. {\bf 82},  403  (1951).

\bibitem{Anderson}
P.~W. Anderson and H. Hasegawa, Phys. Rev. {\bf 100},  675  (1955).

\bibitem{Furukawa2}
N. Furukawa, J. Phys. Soc. Jpn. {\bf 63},  3214  (1994).

\bibitem{Millis}
A.~J. Millis, B.~I. Shraiman, and R. Mueller, Phys. Rev. Lett. {\bf 77},  175
  (1996).

\bibitem{shimomura}
S. Shimomura, N. Wakabayashi, H. Kuwahara, and Y. Tokura, Phys. Rev. Lett. {\bf
  83},  4389  (1999).

\bibitem{Kiryukhin}
V. Kiryukhin, T.~Y. Koo, A. Borissov, Y.~J. Kim, C.~S. Nelson, J.~P. Hill, D.
  Gibbs, and S.-W. Cheong, Phys. Rev. B {\bf 65},  094421  (2002).

\bibitem{Dai}
P. Dai, J.~A. Fernandez-Baca, N. Wakabayashi, E.~W. Plummer, Y. Tomioka, and Y. Tokura,
  Phys. Rev. Lett. {\bf 85},  2553  (2000).

\bibitem{Koo}
T.~Y. Koo, V. Kiryukhin, P.~A. Sharma, J.~P. Hill, and S.-W. Cheong, Phys. Rev.
  B {\bf 64},  220405(R)  (2001).

\bibitem{Lanzara}
A. Lanzara, N.~L. Saini, M. Brunelli, F. Natali, A. Bianconi, P.~G. Radaelli, and
  S.-W. Cheong, Phys. Rev. Lett. {\bf 81},  878  (1998).

\bibitem{Booth}
C.~H. Booth, F. Bridges, G.~H. Kwei, J.~M. Lawrence, A.~L. Cornelius, and J.~J. Neumeier,
  Phys. Rev. Lett. {\bf 80},  853  (1998).

\bibitem{Louca}
D. Louca, T. Egami, E.~L. Brosha, H. R\"{o}der, and A.~R. Bishop, Phys. Rev. B {\bf
  56},  R8475  (1997).

\bibitem{Furukawa}
N. Furukawa, J. Phys. Soc. Jpn. {\bf 66},  2523  (1997).

\bibitem{Schulte}
K. Schulte, M.~A. James, L.~H. Tjeng, P.~G. Steeneken, G.~A. Sawatzky, R.
  Suryanarayanan, G. Dhalenne, and A. Revcolevschi, Phys. Rev. B {\bf 64},
  134428  (2001).

\bibitem{Kuwahara}
H. Kuwahara, Y. Tomioka, Y. Moritomo, and Y. Tokura, Science {\bf 270},  961
  (1995).

\bibitem{Ebata}
K. Ebata, H. Wadati, M. Takizawa, A. Fujimori, A. Chikamatsu, H. Kumigashira,
  M. Oshima, Y. Tomioka, and Y. Tokura, Phys. Rev. B {\bf 74},  064419  (2006).

\bibitem{Saitoh}
T. Saitoh, A. Sekiyama, K. Kobayashi, T. Mizokawa, A. Fujimori, D.~D. Sarma, Y.
  Takeda, and M. Takano, Phys. Rev. B {\bf 56},  8836  (1997).

\bibitem{Mannella}
N. Mannella, A. Rosenhahn, C.~H. Booth, S. Marchesini, B.~S. Mun, S.-H. Yang,
  K. Ibrahim, Y. Tomioka, and C.~S. Fadley, Phys. Rev. Lett. {\bf 92},  166401
  (2004).

\bibitem{Ishida}
Y. Ishida, H. Ohta, A. Fujimori, and H. Hosono, J. Phys. Soc. Jpn.  in press  .

\bibitem{Chaikin}
P.~M. Chaikin and G. Beni, Phys. Rev. B {\bf 13},  647  (1976).

\bibitem{Koshibae}
W. Koshibae, K. Tsutsui, and S. Maekawa, Phys. Rev. B {\bf 62},  6869  (2000).

\bibitem{Wu}
D. Wu, Z.~H. Xiong, X.~G. Li, Z.~V. Vardeny, and J. Shi, Phys. Rev. Lett. {\bf
  95},  016802  (2005).

\end{thebibliography}
\end{document}